\def\jms{J. Mol.\ Spectrosc.\ }

\documentclass[aps,prl,twocolumn,superscriptaddress]{revtex4}
\usepackage{graphics}
\usepackage{verbatim}

\begin{document}

\title{Sub-Doppler frequency metrology in HD for test of fundamental physics}

\author{F. M. J. Cozijn}
\affiliation{Department of Physics and Astronomy, LaserLaB, Vrije Universiteit Amsterdam, de Boelelaan 1081, 1081 HV Amsterdam, The Netherlands}
\author{P. Dupr\'e}
\affiliation{Laboratoire de Physico-Chimie de l'Atmosph\`{e}re, Universit\'{e} du Littoral C\^{o}te d{'}Opale, 189A Avenue Maurice Schumann,
59140 Dunkerque, France}
\author{E. J. Salumbides}
\affiliation{Department of Physics and Astronomy, LaserLaB, Vrije Universiteit Amsterdam, de Boelelaan 1081, 1081 HV Amsterdam, The Netherlands}
\author{K. S. E. Eikema}
\affiliation{Department of Physics and Astronomy, LaserLaB, Vrije Universiteit Amsterdam, de Boelelaan 1081, 1081 HV Amsterdam, The Netherlands}
\author{W. Ubachs}
\affiliation{Department of Physics and Astronomy, LaserLaB, Vrije Universiteit Amsterdam, de Boelelaan 1081, 1081 HV Amsterdam, The Netherlands}
\date{\today}

\begin{abstract}
Weak transitions in the (2,0) overtone band of the HD molecule at $\lambda = 1.38 \, \mu$m were measured in saturated absorption using the technique of noise-immune cavity-enhanced optical heterodyne molecular spectroscopy.
Narrow Doppler-free lines were interrogated with a spectroscopy laser locked to a frequency comb laser referenced to an atomic clock to yield transition frequencies [R(1) = $217\,105\,181\,895\,(20)$ kHz; R(2) = $219\,042\,856\,621\,(28)$ kHz; R(3) = $220\,704\,304\,951\,(28)$ kHz] at three orders of magnitude improved accuracy.
These benchmark values provide a test of QED in the smallest neutral molecule, and open up an avenue to resolve the proton radius puzzle, as well as constrain putative fifth forces and extra dimensions.
\end{abstract}


\maketitle

Molecular hydrogen, the smallest neutral molecule, has evolved into a benchmark quantum test system for fundamental physics now that highly accurate measurements challenge the most accurate theoretical calculations including relativity and quantum electrodynamics (QED) \cite{Piszczatowski2009,Pachucki2016}, even to high orders in the fine structure constant 
(up to $m\alpha^6$) \cite{Puchalski2016}. The measurement of the H$_2$ dissociation energy \cite{Liu2009} was a step in a history of mutually stimulating advancement in both theory and experiment witnessing an improvement over seven orders of magnitude since the advent of quantum mechanics \cite{Sprecher2011}.
Accurate results on the fundamental vibrational splitting  in hydrogen isotopologues \cite{Dickenson2013},
with excellent agreement between experiment and theory, have been exploited to put constraints on the strengths of putative fifth forces in nature \cite{Salumbides2013} and on the compactification of extra dimensions \cite{Salumbides2015b}.

A straightforward strategy to obtain accurate rovibrational level splittings in the hydrogen molecule is to measure weak quadrupole transitions, as was done for H$_2$ in the first \cite{Kassi2014} and second overtone band \cite{Cheng2012,Tan2014}, as well as in the fundamental \cite{Maddaloni2010} and overtone  \cite{Kassi2012,Mondelain2016} bands of D$_2$. In the heteronuclear isotopologue HD, exhibiting a charge asymmetry and a weak dipole moment \cite{Bubin2009b}, a somewhat more intense electric dipole spectrum occurs, first measured by Herzberg \cite{Herzberg1950}. The dipole moment of the (2-0) band is calculated at $20$ $\mu$D  \cite{Pachucki2008}, in reasonable agreement with experiment \cite{Mckellar1973,Kassi2011}.
Accurate Doppler-broadened spectral lines in the HD (2-0) band were reported using sensitive cavity ring down techniques \cite{Kassi2011}. These lines exhibit a width in excess of 1 GHz at room temperature, which challenges the determination of centre frequencies in view of various speed-dependent collisional broadening and shifting phenomena \cite{Wcislo2016}. Careful line shape analysis has led to accuracies of $\sim 30$ MHz, in accordance with the \emph{ab initio} calculated values \cite{Pachucki2010b}.

\begin{figure}
\resizebox{0.45\textwidth}{!}{\includegraphics{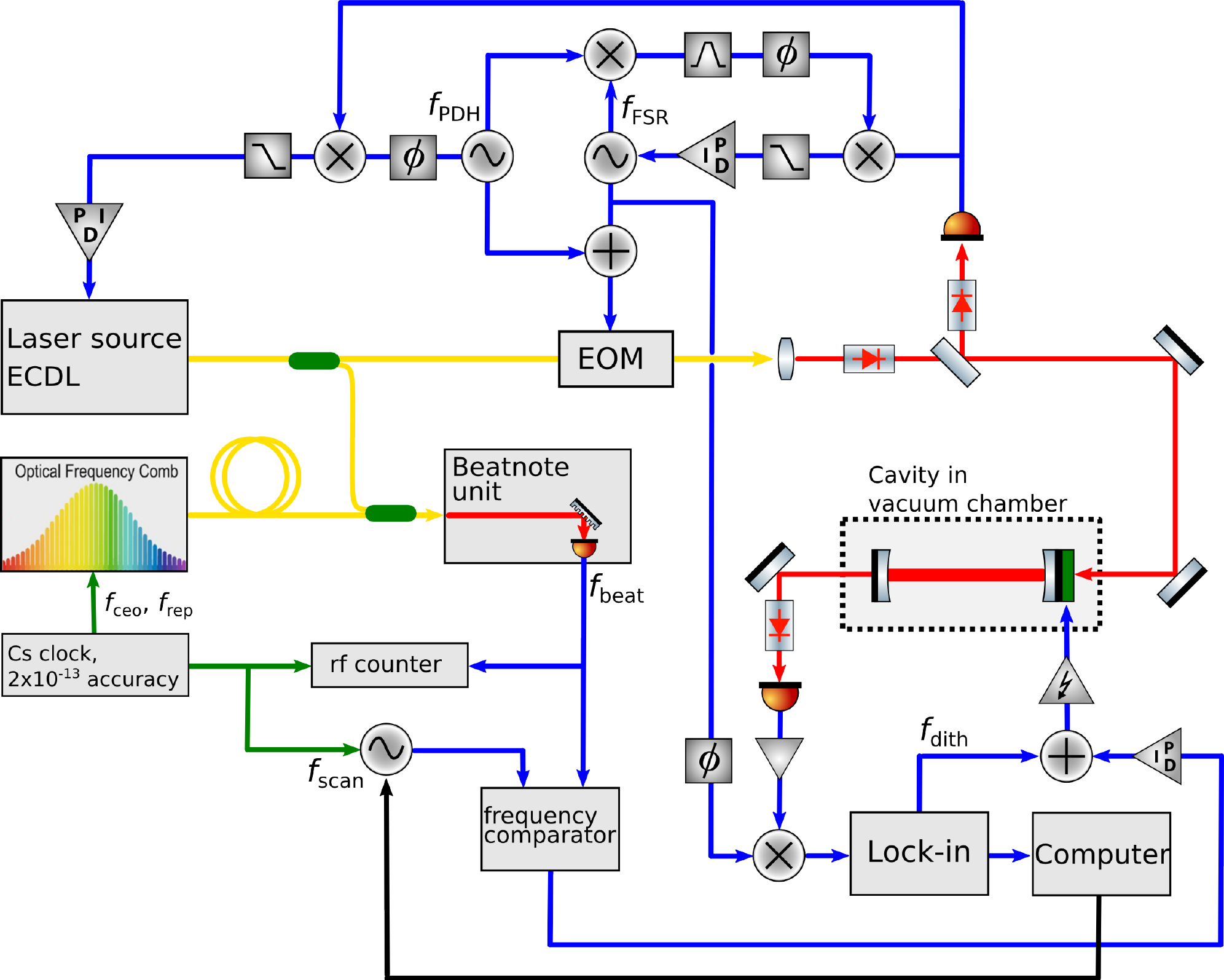}}
\caption{
\small
(Color online) Experimental setup. The spectroscopy laser (ECDL) is sent through a modulator (EOM) to impose both $f_\mathrm{PDH}$ and $f_\mathrm{FSR}$ modulations. $f_\mathrm{PDH}$ is used to stabilize the laser (carrier) frequency to the optical cavity (also the HD absorption cell) and $f_\mathrm{FSR}$ to generate sideband frequencies that are resonant to adjacent cavity modes. The spectroscopy laser is locked to a Cs atomic clock via an optical frequency comb laser for long-term stabilization. Additional cavity-length dither modulation $f_\mathrm{dith}$ is applied for lock-in detection of the HD saturated absorption signals.
}\label{Exp}
\end{figure}

Here we report on the implementation of an absorption technique that combines the advantages of frequency modulation spectroscopy for noise reduction and cavity-enhanced spectroscopy for increasing the interaction length between the light beam and the sample. This extremely sensitive technique, known as  Noise-Immune Cavity-Enhanced Optical Heterodyne Molecular Spectroscopy (NICE-OHMS)~\cite{Ye1996,Ma1999,Schmidt2010,Foltynowicz2011c}, was applied to molecular frequency standards~\cite{Dinesan2015} and to precision measurements on molecules of astrophysical interest~\cite{Markus2016}.
In the present study, weak electric dipole transitions in HD have been saturated, allowing for a reduction of linewidth down to 150 kHz (FWHM), some four orders of magnitude narrower than the Doppler-broadened lines previously reported~\cite{Kassi2011}. The experimental scheme is depicted in Fig.~\ref{Exp}, where the spectroscopy laser is simultaneously locked to the stable optical cavity, and also to a Cs-clock-referenced frequency comb laser to provide an absolute frequency scale during the measurements.

The $48.2$-cm long high-finesse (finesse $\sim 130\,000$)  cavity comprises a pair of curved high-reflectors (Layertec, 1-m radius of curvature), with one of the mirrors mounted on a piezoelectric actuator.
This stabilized optical cavity also provides short-term frequency stability to our spectroscopy laser, and transfers the absolute accuracy of the frequency standard.
The setup provides an intracavity power in the order of 100 W that is sufficient for saturation of HD, while the equivalent absorption path length amounts to $\sim 40$ km.
The cavity itself is enclosed within a vacuum chamber, which can be pumped and filled with the HD gas sample. 

\begin{figure}
\resizebox{0.5\textwidth}{!}{\includegraphics{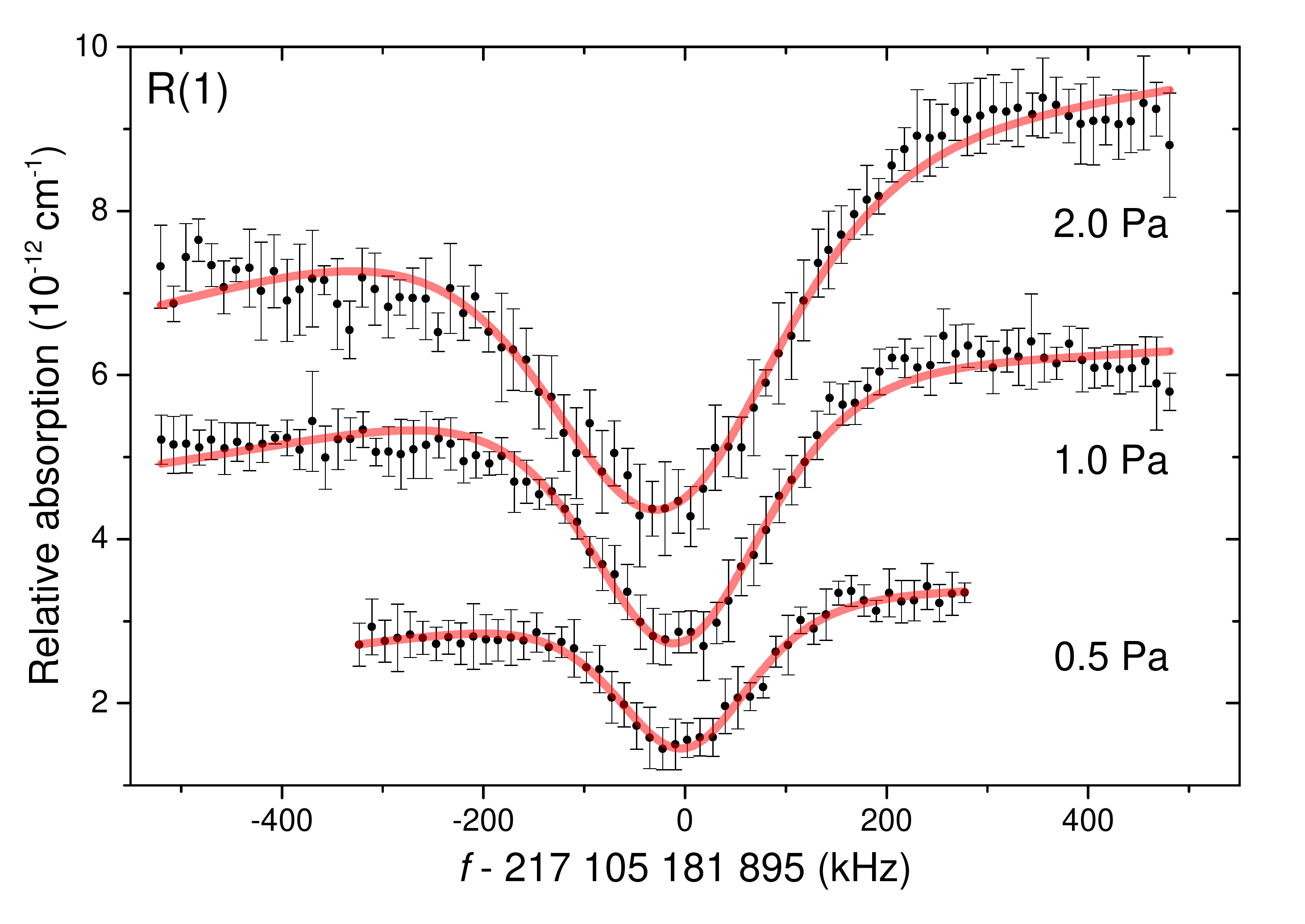}}
\caption{(Color online) Recordings of the HD (2,0) R(1) line for three different pressure conditions averaging 5 scans for 2.0 Pa, 7 scans for 1.0 Pa, and 4 scans for 0.5 Pa.
The solid (red) lines are fits using a line shape function based on a derivative of dispersion~\cite{Ma1999} while allowing for a baseline slope.
The curves have been shifted in the vertical direction for clarity. }\label{R1spec}
\end{figure}

The laser source (ECDL, Toptica DL Pro) operating around $1.38 \, \mu$m is mode-matched and phase-locked to the optical cavity .
The laser beam is fiber coupled and split, with one part for the frequency calibration and metrology, while the main part is phase-modulated through a fiber-coupled EOM (Jenoptik PM1310), allowing for the simultaneous modulation of two frequencies $f_\mathrm{PDH} \sim 20$ MHz and $f_\mathrm{FSR} \sim 310$ MHz.
The reflected beam of the cavity is collected onto an amplified photoreceiver, the signal of which
is used for locking both the laser frequency $f_{\rm{opt}}$ through the Pound-Drever-Hall (PDH) scheme~\cite{Drever1983} and the cavity free spectral range frequency $f_\mathrm{FSR}$ with the DeVoe-Brewer scheme~\cite{DeVoe1984}.
The beam transmitted through the cavity is collected with another high-speed photoreceiver, with the amplified signal demodulated by $f_\mathrm{FSR}$ in a double-balanced mixer.
The resulting dispersive NICE-OHMS signal is sent to a lock-in amplifier to extract the $1f$ signal component at the 
dither frequency $f_\mathrm{dith}\sim 430$ Hz.
The noise equivalent absorption for the setup is estimated to be $3 \times 10^{-13}/(\mathrm{cm}\sqrt{\mathrm{Hz}})$.

\begin{figure}
\resizebox{0.48\textwidth}{!}{\includegraphics{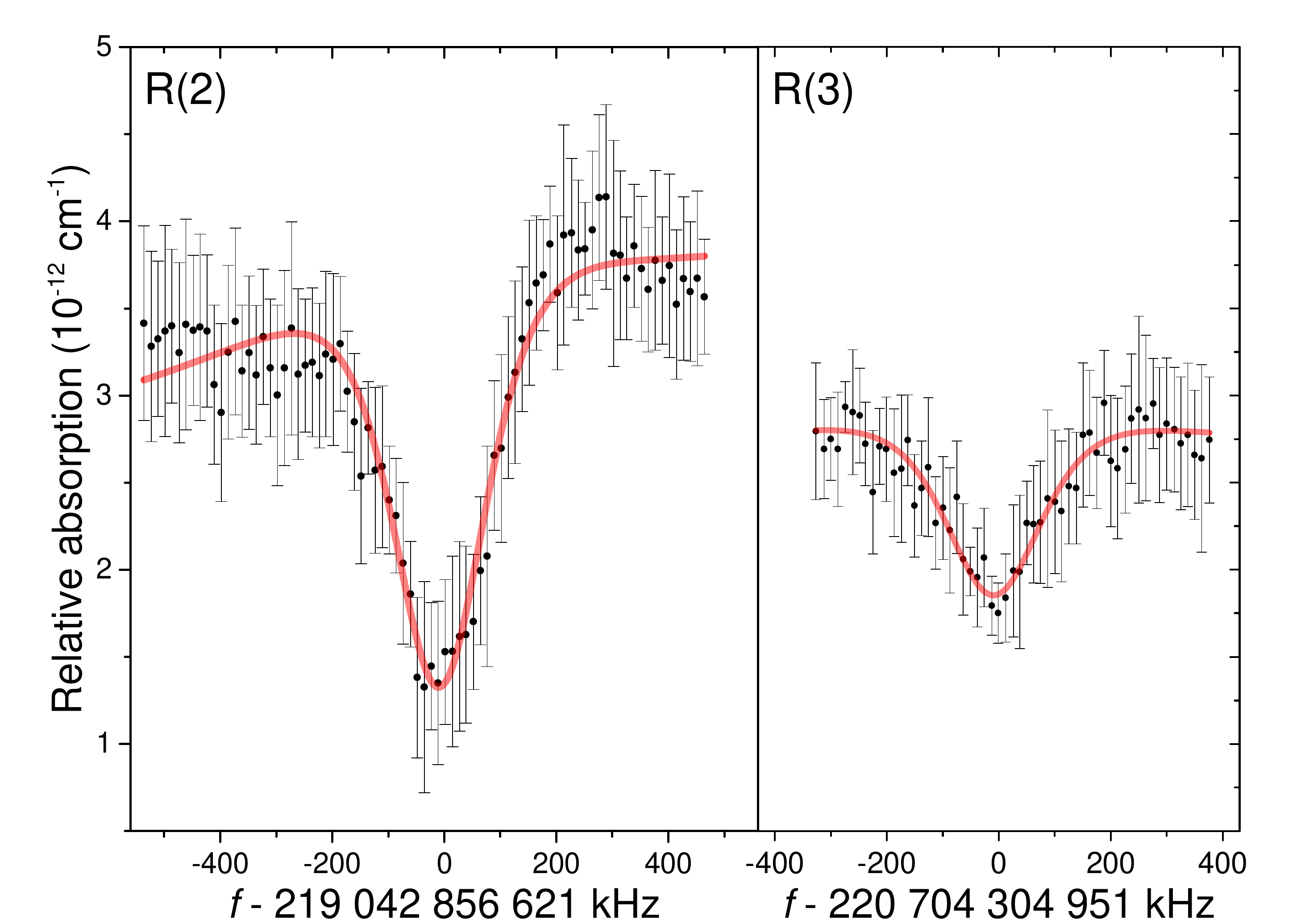}}
\caption{(Color online) Saturation spectra of the R(2) and R(3) transitions of the HD (2,0) overtone band at 1 Pa pressure. [R(2): 12-scan average; R(3): 5-scan average]}\label{R2R3spec}
\end{figure}

The long term frequency stability and accuracy of the system is obtained by beating the spectroscopy laser with a frequency comb (Menlo Systems FC1500-250-WG) stabilized to a Cs clock frequency standard (Microsemi CSIII Model 4301B).
The acquired beatnote frequency $f_\mathrm{beat}$ is counted using an RF counter, and is also used to generate the steering signal for locking the cavity length, thereby tuning the laser frequency $f_\mathrm{opt}$, which is determined via:
\begin{equation}
f_\mathrm{opt} = f_\mathrm{ceo} + n\times f_\mathrm{rep} + f_\mathrm{beat},
\end{equation}
where $f_\mathrm{ceo}=20$ MHz is the carrier-envelope frequency offset of the frequency comb laser, $f_\mathrm{rep}\sim 250$ MHz is its repetition rate, and $n\sim 8.7 \times 10^{5}$ is the mode number.
The absolute frequency of $f_\mathrm{opt}$ is determined with an accuracy better than 1 kHz.

R(1) transitions, recorded at different pressures, are plotted in Fig.~\ref{R1spec}, where each curve is an average of 4 to 7 measurements.
A typical scan takes about 12 minutes, with frequency intervals of 12.5 kHz, and with each data point averaged over 6 seconds.
Fig.~\ref{R2R3spec} displays weaker resonances, where the R(2) spectrum is an average of $12$ scans and that of R(3) an average of $5$ scans.

\begin{figure}
\resizebox{0.48\textwidth}{!}{\includegraphics{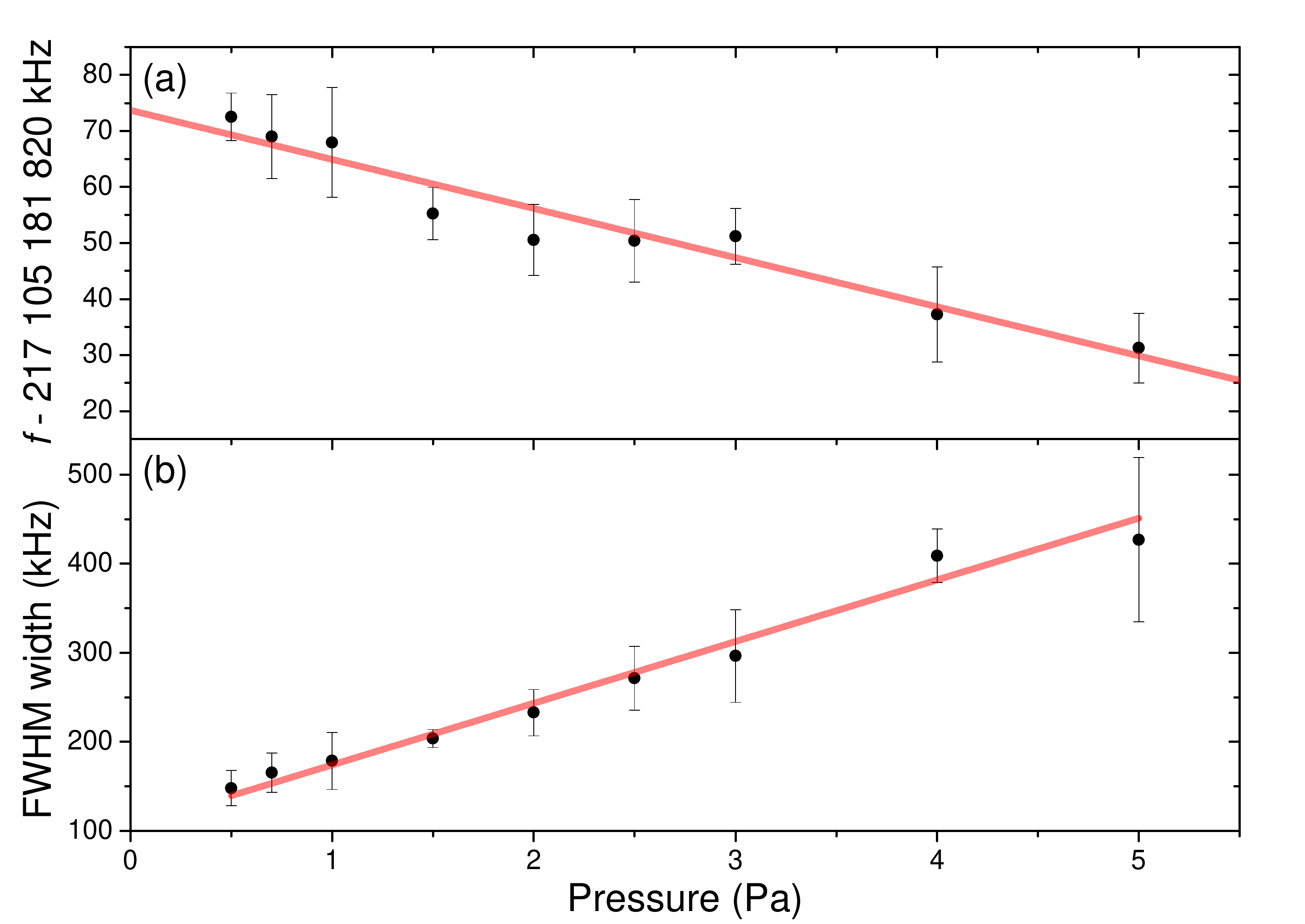}}
\caption{(Color online) Pressure-dependent frequency shift (a) and broadening (b) of the R(1) transition in the $0.5$ to $5$ Pa pressure range.}\label{Pressure}
\end{figure}

The assessment of systematic effects was performed primarily on the R(1) transition, where the signal-to-noise ratio is the highest.
The R(1) transition frequency was measured at different pressures in the range  $0.5-5.0$ Pa (some shown in Fig.~\ref{R1spec}) displaying widths in the range $150-450$ kHz.
This allowed the determination of a pressure-dependent shift coefficient at $-9(3)$ kHz/Pa (see Fig.~\ref{Pressure}(a)) and for extrapolating to a collisionless or zero-pressure transition frequency for R(1).
This collisional shift coefficient is an order-of-magnitude larger (but with similar sign) compared to coefficients for H$_2$ obtained from studies (e.g.~\cite{Cheng2012}) involving pressures higher than kPa.
For R(2) and R(3) transitions measured at 1 Pa, a pressure shift correction of $-9(6)$ kHz was applied.  
This seems appropriate in view of the study on H$_2$~\cite{Tan2014} and D$_2$~\cite{Kassi2012}, where it was shown that the collisional shift parameters only slightly depend on rotational quantum number.

As seen in Fig.~\ref{R1spec}, there is an increase in line shape asymmetry with increasing pressure to which several effects, associated with line broadening (see below), can contribute. In addition, we observe that water vapor absorption in the vicinity of the HD resonances is a likely cause of asymmetry.
This asymmetry ultimately limits the present determination of the transition center to an accuracy of $\sim 1/5$ of the observed resonance width.
 We adopt a phenomenological approach to deal with the asymmetry by applying different line shape profiles, such as Gaussian or Lorentzian, and a function based on a derivative of dispersion~\cite{Ma1999} (plotted in Figs.~\ref{R1spec} and~\ref{R2R3spec}), as well as different baseline fits.
The baseline variation from scan-to-scan can be attributed to residual amplitude modulation.
For the R(1) line, all fits converge to a transition center within $15$ kHz, while a convergence to $20$ kHz is found for the weaker transitions.

\begin{table}
\begin{center}
\caption{
List of corrections $\Delta f$ and uncertainty estimates $\sigma_f$ in units of kHz for the transition frequencies.
}\label{Unc_table}
\begin{small}
\begin{tabular}{l@{\hspace{16pt}}r@{\hspace{16pt}}rc@{\hspace{20pt}}r@{\hspace{16pt}}r}
\noalign{\smallskip}\hline\noalign{\smallskip}
	&\multicolumn{2}{c}{R(1)}&&\multicolumn{2}{c}{R(2), R(3)}\\
\noalign{\smallskip}\cline{2-3}\cline{5-6}\noalign{\smallskip}
Contribution & $\Delta f$ & $\sigma_f$ && $\Delta f$ & $\sigma_f$\\
\noalign{\smallskip}\hline\noalign{\smallskip}
line fitting		& $0$	& 	$15$	&& $0$	& 	$20$\\
pressure shift\footnote{R(1) has been extrapolated to zero pressure, while for R(2) and R(3) a correction is applied based on pressure-shift coefficient of R(1).}
			& $0$	& 	$3$	&& $-9$	& 	$6$\\
2nd-order Doppler 	& $1$	&	$1$	&& $1$	&	$1$\\
ac-Stark shift		& $0$	& 	$10$	&& $0$	& 	$10$\\
frequency calibration	& $0$	& 	$<1$	&& $0$	& 	$<1$\\
\noalign{\smallskip}
subtotal systematic	& $1$	& 	$19$	&& $-8$	& 	$23$\\
statistics		& $0$	&	$10$	&& $0$	&	$15$\\
\noalign{\smallskip}
total			& $1$	& 	$20$	&& $-8$	& 	$28$\\
\noalign{\smallskip}\hline\noalign{\smallskip}
\end{tabular}
\end{small}
\end{center}
\end{table}

Saturation spectroscopy in a cavity leads to a photon recoil doublet that is symmetric to the recoil-free transition center~\cite{Hall1976b}.
For the HD (2,0) transitions at 1.38 $\mu$m, the recoil shift is 34 kHz, that results in a doublet splitting of 68 kHz but does not produce a shift.
At half the intracavity laser power, no significant shift of the line center is observed, and we estimate an upper limit of 10 kHz for the power-dependent or ac-Stark shift.
The second-order Doppler shift is calculated to be $1.2$ kHz for an estimated effective temperature of 185 K. With regard to underlying hyperfine structure  the center-of-gravity of the line will not shift in first order.  For the latter two issues see discussion below.

\begin{table*}
\begin{center}
\caption{
Comparison of R-branch transition frequencies in the HD (2,0) band obtained from the present study with previous experimental determination  $\Delta_\mathrm{exp}$~\cite{Kassi2011}, and with most accurate \emph{ab initio} calculations $\Delta_\mathrm{calc}$~\cite{Pachucki2010b}. Values are given in MHz with uncertainties in units of the last digit indicated in between parentheses. See text for a discussion of the theoretical uncertainty.
}\label{Results}
\begin{small}
\begin{tabular}{c@{\hspace{20pt}}c@{\hspace{20pt}}c@{\hspace{20pt}}c@{\hspace{20pt}}c@{\hspace{20pt}}c}
\noalign{\smallskip}\hline\noalign{\smallskip}
Line & This study & Ref.~\cite{Kassi2011} & $\Delta_\mathrm{exp}$ &Theory~\cite{Pachucki2010b} & $\Delta_\mathrm{calc}$\\
\noalign{\smallskip}\hline\noalign{\smallskip}
R(1)	& $217\,105\,181.895\,(20)$	&	$217\,105\,192\,(30)$	&	$-10$ &	$217\,105\,180$ & $2$\\
R(2)	& $219\,042\,856.621\,(28)$	&	$219\,042\,877\,(30)$	&	$-20$ &	$219\,042\,856$ & $1$\\
R(3)	& $220\,704\,304.951\,(28)$	&	$220\,704\,321\,(30)$	&	$-16$ &	$220\,704\,303$ & $2$\\
\noalign{\smallskip}\hline\noalign{\smallskip}
\end{tabular}
\end{small}
\end{center}
\end{table*}

The collisional or pressure broadening, plotted in Fig.~\ref{Pressure}(b) for the R(1) line, also follows a linear
behavior with a slope  of $70(7)$ kHz/Pa. It is remarkable that the linear trend extends even to the lowest pressure of $0.5$ Pa at which the width is 150 kHz (FWHM). The recoil doublet splitting of 68 kHz and a Rabi frequency of 80 kHz~\cite{Dupre2015a,Dupre2017} must contribute to this width, leaving little room for other contributing phenomena.
The width of $150$ kHz is in itself about ten times narrower than the transit-time limited (FWHM) linewidth of $1.4$ MHz that is expected for HD molecules at room temperature and for the laser beam waist radius of 450 $\mu$m~\cite{Dupre2015a,Dupre2017}.
Similar observations of strongly reduced linewidths below the transit-time limit have been shown in methane~\cite{Bagaev1976,Bagaev1987} and acetylene~\cite{Ma1999}, where it was attributed to the dominant contribution of slow-moving molecules in the saturation signal. Even if the entire width of 150 kHz would be attributed to transit-time broadening an effective kinetic temperature of 185 K would result, but in view of other linewidth contributions the temperature must be significantly lower.
From this we also deduce that hyperfine structure only can have a minor contribution to the line broadening, even though
hyperfine splittings between $F$-components in the HD $(v=0, J=1)$ level span about $220$ kHz~\cite{Ramsey1971}. This may be explained by the hyperfine components in the near-infrared transition overlapping in view of hyperfine level splittings in $v=2$ to be similar as those in $v=0$.

Table~\ref{Unc_table} lists the error budget of the present study.
The statistics entry demonstrates the reproducibility of measurements performed on different days, with the best statistics at $10$ kHz obtained for R(1).
We estimate a total uncertainty, including systematics, of $\sigma_f=20$ kHz for the R(1) transition frequency, and $\sigma_f=28$ kHz for the R(2) and R(3) resonances.
Resulting transition frequencies of the R(1), R(2), and R(3) lines are listed in Table~\ref{Results}.
These values are compared to results of the previous experimental determination by Kassi and Campargue~\cite{Kassi2011} obtained under Doppler-broadened conditions,
showing good agreement, with the present results representing a three order of magnitude improvement in accuracy.
Theoretical level energy calculations by Pachucki and Komasa~\cite{Pachucki2010b}  were claimed to be accurate to $30$ MHz, but values were provided to $3$ MHz ($10^{-4}$ cm$^{-1}$) accuracy.
Since we compare with the energy splittings between $v=0\rightarrow 2$, the theoretical transition frequencies in Table~\ref{Results} should be more accurate due cancellations in various energy contributions.
This assessment of the calculation uncertainty is supported by the excellent agreement between our measurements and the theoretical values that is better than 2 MHz.
%


The $30$-kHz absolute accuracy ($10^{-10}$ relative accuracy) achieved in this study constitutes a thousand-fold improvement over previous work and demonstrates the first sub-Doppler determination of pure ground state transitions in HD, and in fact in any molecular hydrogen isotopologue.
The experimental results challenge current activities in first principles relativistic and QED calculations of the benchmark hydrogen molecules~\cite{Pachucki2010b,Pachucki2016,Pachucki2014,Puchalski2016}.
When such calculations reach the same accuracy level as the experiment there is a potential to constrain theories of physics beyond the Standard Model, as was shown previously \cite{Salumbides2013,Salumbides2015b}.
The finite size of the proton contributes $\sim 300$ kHz to the H$_2$ (3,0) overtone transition~\cite{Puchalski2017} and a similar contribution is expected for the HD transitions investigated here.
If theory and experiment reach the kHz accuracy level, this will allow for a determination of the proton size to 1\% accuracy. Along with complementary investigations in the electronic~\cite{Beyer2017c} and muonic hydrogen atoms~\cite{Pohl2010}, neutral and ionic molecular hydrogen~\cite{Biesheuvel2016}, the HD overtone determinations may contribute towards the resolution of the conundrum known as the proton-size puzzle.

WU acknowledges the European Research Council for an ERC-Advanced grant under the European Union's Horizon 2020 research and innovation programme (grant agreement No 670168).


\end{document}